\documentclass[review]{elsarticle}

\makeatletter
\def\ps@pprintTitle{%
 \let\@oddhead\@empty
 \let\@evenhead\@empty
 \def\@oddfoot{}%
 \let\@evenfoot\@oddfoot}
\makeatother

\usepackage{hyperref}
\usepackage{url}  
\usepackage{ifthen}  
\usepackage{multicol}   
\usepackage[utf8]{inputenc} 
\usepackage{amsmath, amsfonts, amssymb}

\begin{document}

\begin{frontmatter}

\title{An anthropomorphic multimodality (CT/MRI) head phantom prototype for end-to-end tests in ion radiotherapy}


\author[mymainaddress,mysecondaryaddress]{Raya R. Gallas}
\author[mymainaddress,mysecondaryaddress]{Nora H\"{u}nemohr}

\author[mymainaddress,mysecondaryaddress]{Armin Runz}

\author[mymainaddress,mysecondaryaddress]{Nina I. Niebuhr}

\author[mymainaddress,mysecondaryaddress,mytertadd,myquadadd]{Oliver J\"{a}kel}

\author[mymainaddress,mysecondaryaddress]{Steffen Greilich\corref{mycorrespondingauthor}}
\cortext[mycorrespondingauthor]{Corresponding author}
\ead{s.greilich@dkfz.de}

\address[mymainaddress]{German Cancer Research Center (DKFZ), Division of Medical Physics in Radiation Oncology, Im Neuenheimer Feld 280, 69120 Heidelberg, Germany}
\address[mysecondaryaddress]{German Cancer Consortium (DKTK), National Center for Radiation Research in Oncology, Heidelberg Institute of Radiation
Oncology, INF450/400, Heidelberg Germany}
\address[mytertadd]{Heidelberg University Hospital, Department of Radiation Oncology, Im Neuenheimer Feld 400, 69120 Heidelberg, Germany}
\address[myquadadd]{Heidelberg Ion-Beam Therapy Center (HIT), Im Neuenheimer Feld 450, 69120 Heidelberg, Germany}

\begin{abstract}
With the increasing complexity of external beam therapy ``end-to-end'' tests are intended to cover every step from therapy planning through to follow-up in order to fulfill the higher demands on quality assurance. As magnetic resonance imaging (MRI) has become an important part of the treatment process, established phantoms such as the Alderson head cannot fully be used for those tests and novel phantoms have to be developed. Here, we present a feasibility study of a customizable multimodality head phantom. It is initially intended for ion radiotherapy but may also be used in photon therapy.\\ 
As basis for the anthropomorphic head shape we have used a set of patient computed tomography (CT) images. The phantom recipient consisting of epoxy resin was produced by using a 3D printer. It includes a nasal air cavity, a cranial bone surrogate (based on dipotassium phosphate), a brain surrogate (based on agarose gel), and a surrogate for cerebrospinal fluid (based on distilled water). Furthermore, a volume filled with normoxic dosimetric gel mimicked a tumor. \\
The entire workflow of a proton therapy could be successfully applied to the phantom. CT measurements revealed CT numbers agreeing with reference values for all surrogates in the range from 2\,HU to 978\,HU (120\,kV). MRI showed the desired contrasts between the different phantom materials especially in T2-weighted images (except for the bone surrogate). T2-weighted readout of the polymerization gel dosimeter allowed approximate range verification. \\ 
\newline
\textbf{Ein Prototyp f\"{u}r ein anthropomorphes multimodales (CT/MRT) Kopfphantom f\"{u}r End-zu-End-Tests in der Ionenstrahl-Therapie}
\newline
\newline
Die h\"{o}here Pr\"{a}zision der Strahlentherapie erfordert zugleich eine Verbesserung der Qualit\"{a}tssicherung, welche den gesamten Ablauf der Strahlentherapie abdecken muss. Da die Magnetresonanztomographie (MRT) eine zunehmend wichtige Rolle spielt, k\"{o}nnen bisherige Phantome (wie der Alderson Kopf) dabei nur begrenzt eingesetzt werden. Diese Arbeit stellt eine Machbarkeitsstudie f\"{u}r ein multimodales Kopfphantom dar, welches f\"{u}r die Ionenstrahl-Therapie konzipiert ist, aber auch in der Photonentherapie zum Einsatz kommen kann.\\
Die anthropomorphe Kopfform basiert auf dem Computertomographie (CT)-Datensatz eines Patienten und wurde mithilfe eines 3D-Druckers aus Epoxidharz geformt. Bef\"{u}llt wurde das Phantom mit einem Knochensurrogat (Dikalium\-hydrogenphosphat-L\"{o}sung), einem Gehirnsurrogat (Agarosegel) und einem Ventrikelsurrogat (destilliertes Wasser). Enthalten sind des Weiteren eine nasale Luftkavit\"{a}t sowie ein Dosimetriegel als Tumorsurrogat.\\
Das erstellte Phantom konnte den gesamten Prozess einer Bestrahlung mit Protonen durchlaufen. CT-Messungen des Phantoms ergaben mit den Referenzwerten nat\"{u}rlicher Gewebe vergleichbare Hounsfield-Einheiten (2\,HU bis 978\,HU bei 120\,kV). Das Phantom lieferte auch in MRT-Messungen Kontraste, vor allem in T2-gewichteten Bildern. Die Auswertung des Dosimetriegels mithilfe von MRT erm\"{o}glichte eine Reichweitenbestimmung der Protonen.\\ 
\end{abstract}


\end{frontmatter}

\section{Background}
Over the last decade radiation therapy (RT) has achieved significantly higher accuracy in dose delivery due to the improved image guidance and advanced treatment delivery techniques. Consequently, adequate quality assurance becomes increasingly important, with a special focus on the integration of the various techniques involved in the clinical workflow. This affects especially the integration of MRI which is already utilized for target definition and is widely expected to gain even more importance in image guidance. Recently, the German Commission on Radiological Protection (Strahlenschutzkommission) \cite{Bundesjustizministerium2011} recommended to test the entire chain of radiation therapy by system-wide ``end-to-end'' tests. These tests are also claimed by the German Radiation Protection Ordinance (Strahlenschutzverordnung, StrSchV) \cite{Veith2005} while the Standards Committee for Radiology (Normenausschuss Radiologie, NAR) is working out corresponding standards for the coming years.\\
Such tests call for phantoms able to represent the patient in a number of essential radiological aspects, such as the contrasts from different imaging modalities for diagnosis, target delineation, and treatment simulation. Furthermore, phantoms should allow to test different techniques and tools for patient fixation and dose delivery.\\
The Alderson head is most likely the best known example for a phantom providing anthropomorphic structures with realistic contrasts in CT. Being based on bony material and polymers, it cannot however, be used with MR imaging. Additionally, it hardly allows three-dimensional dosimetry and is limited to pointwise dose measurements (by inserting 1D dosimeters such as ion chambers) and two-dimensional measurements (by inserting films into designated slots).\\
It is therefore desirable to develop multimodality phantoms providing not only contrasts in both CT and MRI but also opening the possibility of a more flexible dosimetry. In this article we present a prototype of a phantom which has been based on a head shaped recipient with compartments for surrogates of brain, bone, and ventricles. In the target volume, polymerization gel was placed \cite{baldock2010}. During our study, the phantom underwent CT and MR imaging followed by treatment planning and proton irradiation. Subsequently, range verification was performed by an MRI readout of the polymerization gel dosimeter in the target volume.\\
%
%
%
\section{Material and Methods}
%
\subsection{Phantom Recipient}
\begin{figure}[htbp]
  \centering
  \includegraphics[width=0.9\textwidth]{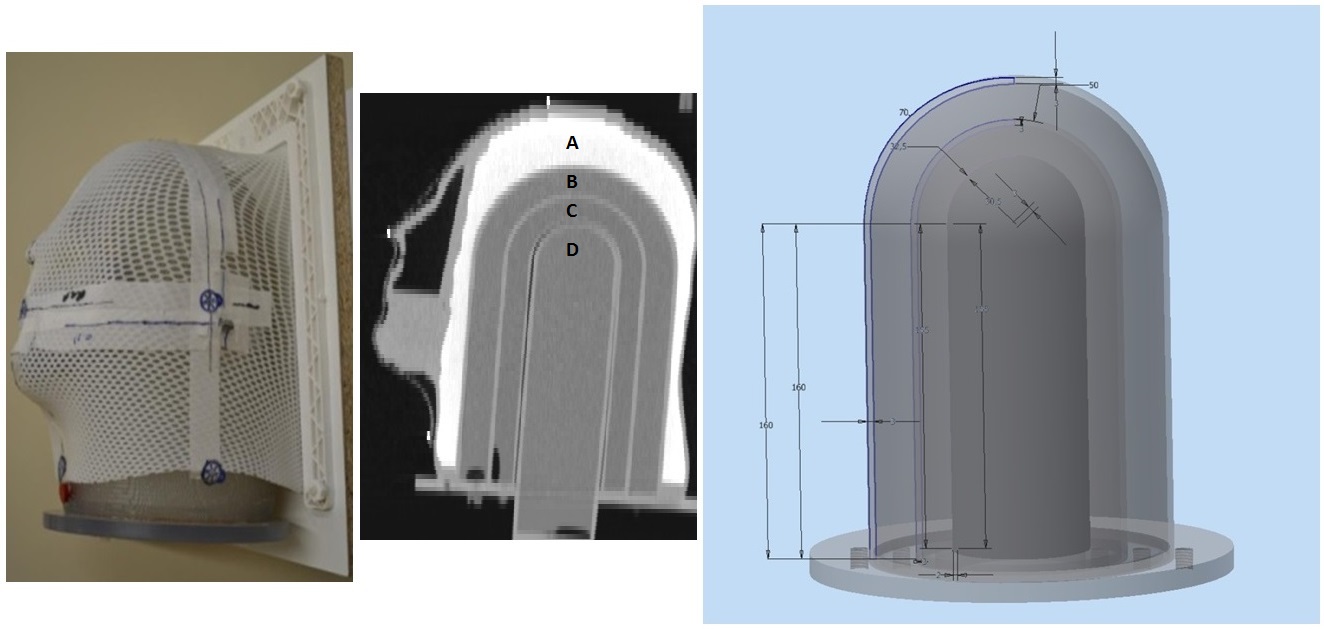}
  \caption{The prototype head phantom. Left part: the phantom fixated with a thermoplastic mask. Middle: CT image cross section of the phantom (A: cranial bone surrogate, B: brain tissue surrogate, C: cerebrospinal fluid surrogate, D:~polymerization gel dosimeter). Right part: inner cylindrical structure set.}
  \label{fig1}
\end{figure}

The phantom recipient was produced by means of 3D printing using epoxy resin as the printing material. 3D printing allows the production of stable objects of almost any shape. One of many possible applications in the future might be the production of individualized phantoms according to patient geometries \cite{Rengieretal2013}. \\
The phantom recipient was composed of two separated parts as a compromise between anatomic reality and geometric shapes that are easy to handle (e.g. for filling with tissue surrogates, Fig.\,\ref{fig1}) without loss of generality for the approach itself. The anthropomorphic outer shape was modelled by segmenting a head outline from a CT-dataset using the MITK software (version 2012.12.00, segmentation tool ``Add'') \cite{Wolfetal2005}. A simplified nasal cavity was added and a solid chin part made from printed epoxy resin. The second part of the phantom consists of a system of inner cylinders constructed using Autodesk Inventor Professional 2013 \cite{autodesk} to house the tissue surrogates and the polymerization gel dosimeter in the target volume. After printing, both parts were combined and nested. Access to the inner compartments was assured by holes in the neck region. Compartment D (Fig.\,\ref{fig1}) was printed separately and can be exchanged easily.\\
%
\subsection{Phantom Materials}
We restricted our study to the following, most essential materials found in the human head.\\
%
\subsubsection{Bone Surrogate}
Compartment A of the phantom (Fig.\,\ref{fig1}) contains a surrogate representing cranial bone using a solution of 1250\,g dipotassium phosphate ($\text{K}_2\text{HPO}_4$) in one liter of distilled water. Solutions of dipotassium phosphate allow the adjustment of electron densities comparable to those of bone for CT. Furthermore, their chemical composition is similar to bony tissues (esp. ``ribs 10th'' from \cite{schneider2000correlation}) which is advantageous for realistic mimicking of photon and ion interaction processes with the material. However, the ratio of carbon and oxygen in polymer surrogates cannot be exactly the same as for tissue, e.g. the bone liquid contains no carbon in contrast to real bone (Tab.\,\ref{tab1}). This is tolerable since the effective atomic number and the I-value of carbon and oxygen are similar. An advantage of the liquid bone surrogate is the easy positioning in the phantom recipient with minimal air inclusions and the easy removal or replacing. Heating the water to $50^\circ\text{C}$ helps to dissolve the elevated concentration of dipotassium phosphate in water.\\
\begin{table}[htbp]
{\tiny
\caption{Density $\rho$, atomic composition, relative electron density to water 
$\rho_{e^-}$, effective atomic number~$Z_{eff}$, and mean excitation potential $I$ of the 
investigated materials. ($^{a}$\cite{Uusi2003})}
  \label{tab1}
 \hspace*{-0.5cm}
\begin{tabular}{l l l l l l l l l l }
\hline
Surrogate	  	& $\rho$ [$\frac{g}{cm^3}$]				& H			
& O   & C		& P		& K   & $\rho_{e^-}$	& $Z_{eff}$ & I [eV] \\
\hline
Bone liquid 	  	& 1.53				& 0.054			& 0.603   & -		
& 0.097		& 0.246   & 1.308	& 12.82 & 104.5 \\
Brain gel/agarose gel  	& 0.96				& 0.109			& 0.024   & 0.867	
	& -		& -   & 0.958	& 7.42 & 75.4 \\
Ventricle/water  	& 1.00				& 0.112			& -   & 0.888		
& -		& -   & 1.000	& 7.45 & 75.3 \\
Soft bone/printing material	& 1.21				& n/a			& n/a   & 
n/a		& n/a		& n/a   & 1.181	& 9.28 & n/a \\
Tumor volume/dosimetric gel 	& 1.05$^{a}$				& 0.104$^{a}$			
& 0.105$^{a}$   & 0.767$^{a}$		& -		& -   & 1.021	& 7.30 & 74.5 \\
\hline
\end{tabular}}
  \end{table}

%
\subsubsection{Brain Surrogate}
\begin{figure}[htbp]
  \centering
  \includegraphics[width=0.9\textwidth]{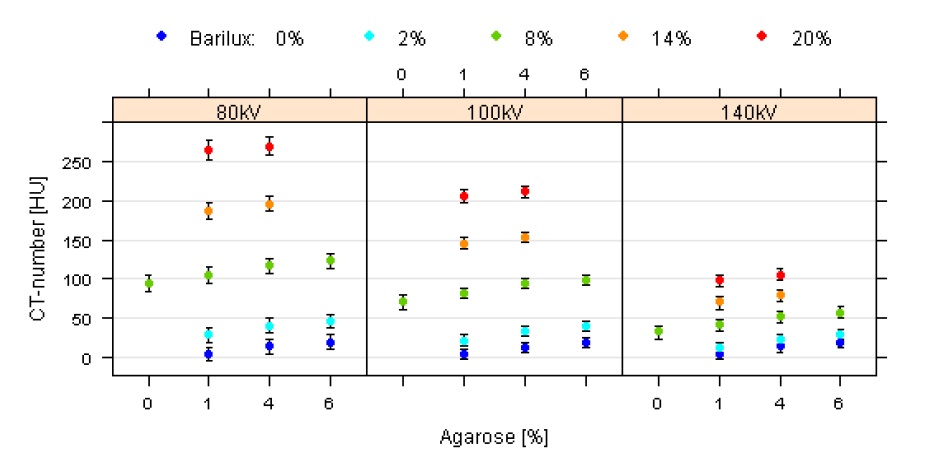}
  \caption{CT-numbers of agarose gels with a concentration up to 6\,\% for a series of 
tube voltages and additional CT contrast agent. Adding a high Z CT contrast agent such as Barilux 
(based on bariumsulfate) opens the possibility to vary CT and MR contrasts independently. CT 
contrast agents hardly influence the T1 and T2 contrast. Vice versa, concentrations of MR contrast 
agents (e.g. gadolinium) of a few  mmol/L would hardly influence CT contrasts.}
  \label{fig2}
\end{figure}

Compartment B (Fig.\,\ref{fig1}) represents the brain tissue and contains a gel made from agarose powder and distilled water. By adjusting the concentration of agarose in gels, T2 and to a smaller extent T1 MR relaxation times can be modified. On the other hand, agarose hardly influences the electron or mass density and consequently the CT contrast will remain largely the same within different agarose gels (Fig.\,\ref{fig2}). Due to the polysaccharide polymer material, agarose also features a tissue equivalent composition and is therefore an adequate surrogate to mimic various soft tissues corresponding to good MRI but poor CT contrasts.
In general, we found that agarose concentrations of 1-6\,\% are feasible to solute. In this study, a brain surrogate was made by heating 105.25\,g of agarose powder in 2000\,g distilled water (corresponding to a concentration of 5\,\% agarose). The mixture from agarose and distilled water needs to boil for at least one minute and can be drawn into the recipient by using a vacuum pump. \\
%
%
\subsubsection{Ventricle Surrogate}
The cerebrospinal fluid in the ventricles is very similar to water. For that reason, distilled water was filled into compartment C (Fig.\,\ref{fig1}) which additionally provides a standard for MR as well as for CT imaging.
%
%
\subsubsection{Tumor Surrogate/Polymerization Gel Dosimeter}
While the previous materials were chosen for achieving realistic contrasts in CT and MR imaging, the polymerization gel dosimeter in compartment~D (Fig.\,\ref{fig1}) mainly served the range verification. For this study, a BANG\textsuperscript{\textregistered}3-Pro\textsuperscript{\texttrademark}Gel Dosimeter Kit bought from MGS was prepared by following the instructions given by the manufacturer. During irradiation, water molecules contained in the gel start dissociating into free radicals which induce the polymerization of monomers present in the gel. This dose dependent polymerization reduces the mobility of free water molecules causing a change in the T2 relaxation time of the irradiated gel which can be detected in 3D by T2-weighted MR acquisitions. Besides the dosimetry purpose, the gel also represents a further soft tissue contrast similar to ICRU~44 brain tissue \cite{Uusi2003}.\\
Soft tissue contrast, the easy incorporation of the dosimeter in the anthropomorphic shape, and the possibility to perform 3D dosimetry were the main reasons for choosing a polymerization gel dosimeter, despite its current limitations for proton and ion dosimetry.
%
%
\subsubsection{Soft Bone/Printing Material}
The printing material forms the outermost layer of the phantom as well as the separation layers between the surrogates and a massive block in the chin region. It is an artificial epoxy resin and being a polymer with a density of 1.21\,$\text{g}/\text{cm}^3$ (cured), it also serves as a soft bone surrogate.\\
%
\section{Experiments}
The head phantom filled with the tissue surrogates was screwed to a wooden board using a thermoplastic mask (Fig.\,\ref{fig1}). Subsequently, the head phantom followed the workflow process of a typical proton radiotherapy treatment course with pre-treatment imaging via CT and MR, treatment planning on CT images, delivery, and range verification by using the MR readout of the polimerization gel dosimeter.\\
%
\subsection{Pre-treatment MR Imaging}
The head phantom was positioned in the 12-channel head coil of a Siemens 1.5\,T MAGNETOM Avanto scanner in order to acquire the pre-treatment MR image. T1- and T2-weighted images (4\,mm slices) were acquired using a flash 2D (TR=110\,ms, TE=4.8\,ms) and a turbo spin echo (TR=3500\,ms, TE=88\,ms) sequence, respectively. The phantom was removed from the wooden board and the immobilization mask to fit into the used head coil. The obtained images were used for analyzing the similarity between the phantom's and real tissues' MR contrasts as well as for the comparison of the polymerization gel dosimeter before and after the irradiation. \\
%
\subsection{Material Characterization in a Dual Energy CT Scanner}
\label{sec:DECT}
\begin{figure}[htbp]
  \centering
  \includegraphics[width=0.6\textwidth]{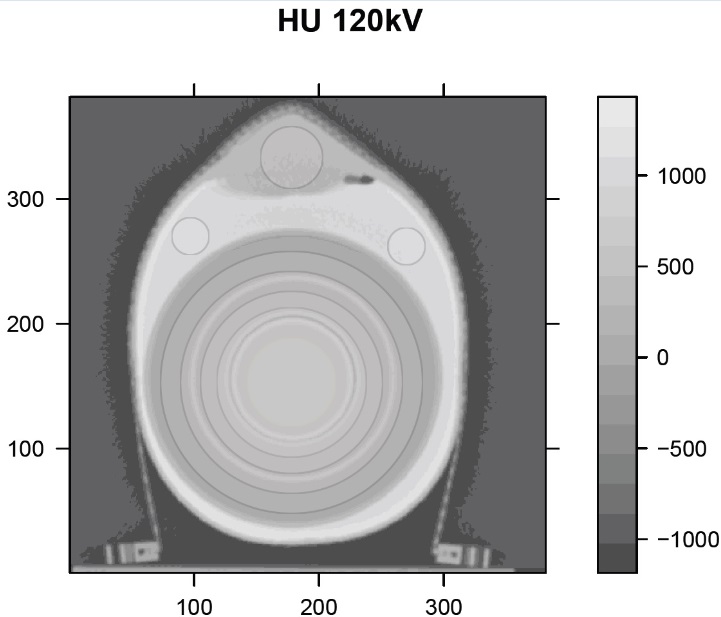}
  \caption{VOI selection in the transversal CT plane to evaluate image values (HU, ED, 
Zeff) prediction of single and dual energy CT for the investigated phantom materials.}
  \label{fig3}
\end{figure}

The immobilized head phantom was imaged in a dual energy CT scanner (Siemens Somatom Definition Flash) at tube voltages of 80\,kVp and 140\,kVp/Sn (higher spectrum is hardened by a tin filter). Tube currents (eff.\,mAs) were set to 220\,mAs\,(80\,kVp)/85\,mAs\,(140\,kVp/Sn) with a default pitch of 0.6 and a rotation time of 1\,s. 3\,mm slice images were reconstructed with a D30 kernel. Both image stacks were further processed by a dedicated algorithm \cite{Hunemohretal2014} to calculate an electron density and effective atomic number image stack. Electron density and effective atomic number were evaluated to explore tissue equivalence. To obtain the CT numbers, the images were processed by ``R'' \cite{RDevelopmentCoreTeam2010} using the ``oro.dicom'' package. Gray values were analyzed in eleven slices using the circular VOIs shown in Fig.\,\ref{fig3}.\\
%
%
\subsection{Tumor Localization and Treatment Planning}
\label{sec:SECT}
\begin{figure}[htbp]
  \centering
  \includegraphics[width=0.6\textwidth]{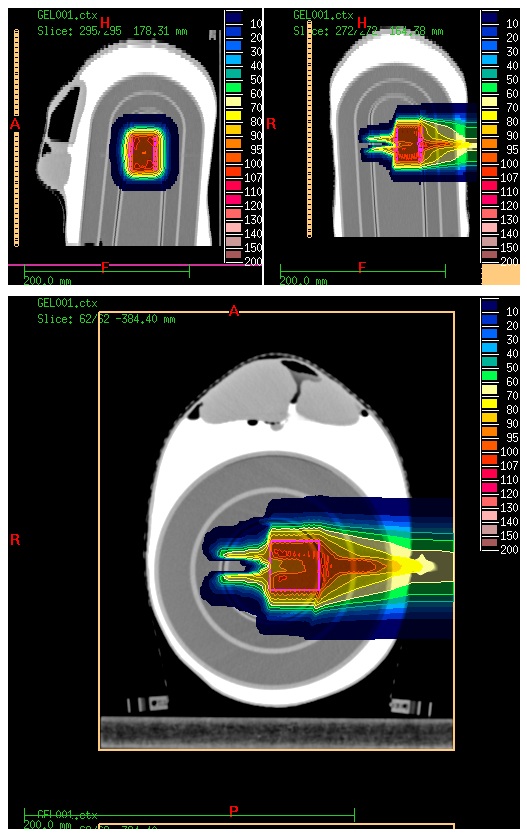}
  \caption{Planning CT including PTV (pink) and isodose lines. An 
overshoot is clearly seen due to a tungsten pin attached to the mask which was used to simulate the 
impact of a dental implant on the dose distribution.}
  \label{fig4}
\end{figure}

CT imaging for treatment planning was acquired with the Siemens Biograph mCT operated in clinical routine for treatment planning at the Heidelberg Ionbeam-Therapy Center (HIT).\footnote{Dual CT used for material characterization (sec.\,\ref{sec:DECT}) only as it is not yet part of clinical routine.} Image acquisition was done with 120\,kVp and the standard planning protocol (eff.\,mAs of 300, 3\,mm slice thickness, H40 reconstruction kernel, 310\,mm FOV). The CT number conversion to relative stopping-power ratio for ions was assured by a standard stoichiometric calibration \cite{Schneideretal1996}.\\
A target volume of 5x5x3.5\,$\text{cm}^3$ was defined for an irradiation at a horizontal beamline at HIT with 2\,Gy physical proton dose. To allow subsequent monitoring of the steep proton dose fall off in the polymerization gel dosimeter, the irradiation was planned such that the dose maximum reached up to the center of the gel (Fig.\,\ref{fig4}). Plan optimization was performed with the treatment planning system syngo\textsuperscript{\textregistered} RT Planning VB10.\\
%
%
\subsection{Phantom Positioning and Treatment}
Crosshairs on the basis of the single energy CT scanner’s lasers were sketched onto the phantom. Because the single energy CT scanner delivered the dataset for treatment planning, the phantom was positioned on the basis of those crosshairs at the horizontal irradiation place. Irradiation was performed at HIT using active raster scanning of protons in 28 iso-energy slices (between 87.53 and 140.97\,MeV corresponding to a range between 6.1 and 14.3\,cm in water) to cover a volume of 5x5x3.5\,$\text{cm}^3$. The dose rate was approximately 3.0 (8.2)\,$\tfrac{Gy}{l \cdot min}$ including (without) spill pauses.\\
\subsection{Range Verification with the MR readable polymerization gel dosimeter}
\begin{figure}[htbp]
  \centering
  \includegraphics[width=0.9\textwidth]{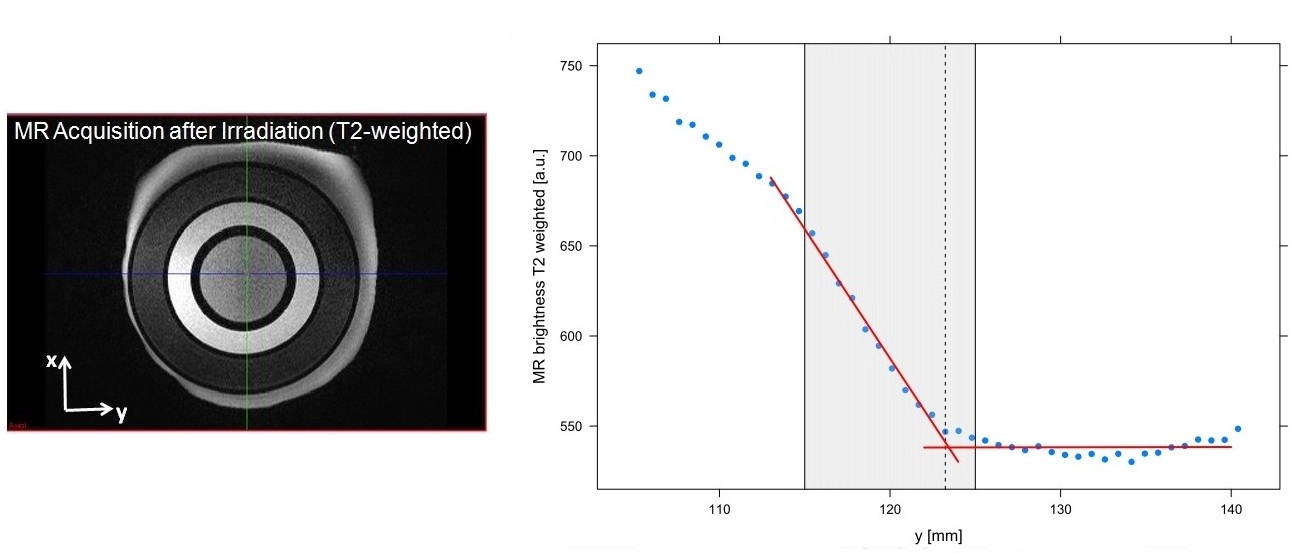}
  \caption{Left part: T2-weighted MR image; the distal beam edge is hardly 
visible (lower signal intensities in the irradiated right part of the polymerization gel dosimeter). 
Right part: mean MR signal profile in the polymerization gel dosimeter along the proton beam axis in 
the area of the SOBP; the distal edge is more obvious as this curve is averaged across x. The 
dashed line is parallel to the distal edge and passes through the central point of the innermost 
compartment. Thus, it represents the planned edge of the dose maximum. The intersection of the 
fitted red lines represents the measured edge of the dose maximum which is positioned about 0.3\,mm 
right of the planned edge.}
  \label{fig5}
\end{figure}

For range verification, the polymerization gel dosimeter in compartment D (Fig.\,\ref{fig1}) was read out using a T2-weighted turbo spin echo sequence (TR=4120\,ms, TE=87\,ms, slice thickness 1.5\,mm, 1.28 pixel/mm). Only the irradiated part of the polymerization gel dosimeter and its immediate surrounding were imaged.\\
The MR acquisitions were saved as a DICOM data set and the profile in Fig.\,\ref{fig5} was obtained by processing the pixel values by the ``R'' software~\cite{RDevelopmentCoreTeam2010}. To do so, the brightness profiles in the irradiated part along the beam axis were superposed and averaged. Two lines were adapted to this average profile and their intersection was determined.\\
We did not perform a registration of the MR acquisitions with the treatment plan but we compared the positions in the gel relative to the central point of compartment D (Fig.\,\ref{fig1}) which was possible since pixel size and compartment size were known.
%
%
\section{Results}
%
\subsection{MR Imaging}
\begin{figure}[htbp]
  \centering
  \includegraphics[width=0.6\textwidth]{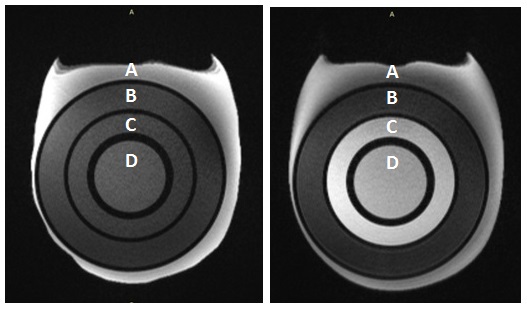}
  \caption{T1- (left) and T2-(right) weighted transversal cross section of the head 
phantom. A: cranial bone surrogate, B: brain tissue surrogate, C: cerebrospinal fluid surrogate, D: 
polymerization gel dosimeter. }
  \label{fig6}
\end{figure}

Fig.\,\ref{fig6} shows a T1- and a T2-weighted transversal cross section of the phantom. Contrasts are more pronounced in T2-weighted acquisitions and materials can be distinguished except for the printing material and air whose signal intensities are both very low. The brain surrogate in compartment B shows a low signal in both T1- and T2-weighting and reacts therewith similar to muscle tissue. The polymerization gel dosimeter in compartment D can be regarded as a soft tissue surrogate and shows a rather high signal in the T2-weighted acquisition and a low signal in the T1-weighted image - this is typical for liquids and water (see also the ventricle surrogate in compartment C). The bone surrogate (A) does not meet the requirements of realistic MR signal intensities as it shows high signal intensity in both weightings due to its aqueous nature (Fig.\,\ref{fig6}). \\
%
%
\subsection{Material Characterization}
CT images of the dual energy CT (see \ref{sec:DECT}) and the single energy planning CT scanner (see \ref{sec:SECT}) were registered with a mutual information algorithm implemented in the planning system ``VIRTUOS'' at DKFZ. Tab.\,\ref{tab2} summarizes the mean and standard deviations of the phantom materials' CT numbers, electron density, and effective atomic number. \\
 \begin{table}[htbp]
 {\tiny
  \caption{Measured CT numbers, relative electron densities and effective atomic numbers.}
 \label{tab2}
  \hspace*{-0.75cm}
\begin{tabular}{l l l l l l}
\hline
Material 	& $\text{CT no}_{80kV} \text{[HU]}$ 	 	& $\text{CT no}_{120kV} \text{[HU]}$ 
	& $\text{CT no}_{140kV} \text{[HU]} $	& $\rho_{e^-}$	& $Z_{eff}$ \\
\hline
Ventricle/water	& $16\pm23$ 	& $2\pm7$  & $5\pm17$ & $1.000\pm0.012$  	& $7.7\pm0.4$  \\
Brain gel/agarose gel 	& $45\pm24$ 	& $24\pm8$  & $24\pm16$ & $1.015\pm0.011$  	& 
$7.9\pm0.3$  \\
Soft bone/printing material 	& $357\pm24$ 	& $286\pm15$  & $243\pm16$ & $1.191\pm0.011$  	& 
$9.4\pm0.2$  \\
Tumor volume/dosimetric gel  	& $52\pm25$ 	& $41\pm7$  & $43\pm18$ & $1.039\pm0.012$  	& 
$7.6\pm0.4$  \\
Bone liquid	& $1298\pm35$ 	& $978\pm24$  & $774\pm18$ & $1.54\pm0.013$  	& $12.2\pm0.1$  \\
\hline
\end{tabular}}
  \end{table}
%
\subsection{Gel read out}
\label{sec:Gel read out}
Fig.\,\ref{fig5} (left part) visualizes the T2-weighted MR acquisitions of the polymerization gel dosimeter taken after irradiation. The brightness profile (Fig.\,\ref{fig5}, right part) shows a significant signal drop from left to right.\\
%
%
\section{Discussion}
%
\subsection{Imaging Contrasts and Material Properties}
CT numbers of the investigated surrogates were found to be comparable to simulated reference values found in \cite{Saitoetal2012} (bone solution to mandible: max.\,deviation 35\,HU; brain surrogate and agarose gel to reference brain tissue: max.\,deviation 13\,HU; printing material to spongiosa: max.\,deviation 34\,HU). Electron densities are consistent to the reference electron densities of the corresponding tissues and deviate by maximum 0.04 (absolute value). Effective atomic numbers agree within 0.2-1.1 (absolute values) compared to reference values \cite{saito2011optimized} (Tab.\,\ref{tab3}).\\
\begin{table}[htbp]
\centering
  \caption{Reference values of three tissues taken from \cite{saito2011optimized}, 
\cite{Saitoetal2012}. These tissues show comparable attributes to the materials investigated in this 
work. Please mind that effective atomic numbers were calculated with a slightly different exponent 
of m=3.5 in \cite{Saitoetal2012} as compared to this work (m=3.1, \cite{Hunemohretal2014}) and are 
therefore larger than measured $Z_{eff}$ in this work.}
  \label{tab3}
\begin{tabular}{l c c c c }
\hline
Material	  	&  $\rho_{e^-}$			& $Z_{eff}$ & $\text{CT no}_{80kV} 
\text{[HU]}$ & $\text{CT no}_{140kV} \text{[HU]} $ \\
\hline
Brain & 1.035 & 7.65 & 43 & 37 \\
Spongiosa & 1.150 & 10.49 & 339 & 209 \\ 
Mandible & 1.577 & 13.16 & 1326 & 809 \\
\hline
\end{tabular}
  \end{table}
MR acquisitions showed signal intensities of the soft tissue surrogates that are qualitatively similar to the natural equivalents’ intensities. Improvements of the phantom could therefore rather focus on more realistic and further differentiated materials for MR imaging. The excellent MR based soft tissue distinction is used for target and normal structure delineation in RT and should be provided in a future ``end-to-end'' test phantom. Additionally, a solid bone surrogate with no MR signal would be desirable to test e.g. matching algorithm. It could be realized for example with gypsum or even with the printing material itself. By doing the latter, a more realistic anatomy would be possible by directly printing the bone matrix of the patient specific CT data plus additional compartments for soft tissue. \\
%
\subsection{Range Verification}
The brightness profile in the polymerization gel dosimeter does not show a sharp distal edge of the spread out Bragg peak (SOBP) (Fig.\,\ref{fig5}, right), which is likely due to partial volume detection \cite{berg2013high}, i.e. the dose in the Bragg-peak region can be misjudged due to averaging over the voxel size. In addition, quenching can occur for high LET protons~\cite{heufelder2003}.\\
Our method to compare the planned and the measured dose edge has to be improved for future applications as it cannot be applied to patients with a considerably complexer geometry; the registration of treatment plan and MR images would therefore be most preferable.\\
Furthermore, the determination of the distal edge from the measured data is subject to an uncertainty of approximately 2\,mm in the position of the compartment center in the MR acquisitions as well as due to procedure of fitting two lines to the profile in Fig.\,\ref{fig5}. In addition, degradation of the distal edge occurs due to the impact of the tungsten pin (compare Fig.\,\ref{fig4}). Nevertheless, the planned (dashed line in Fig.\,\ref{fig5}) and the measured range (intersection of the fitted lines in Fig.\,\ref{fig5}) agree within this uncertainty, stressing that range verification using polymerization gel dosimeters is principally possible. Future studies should also exploit 3D dose readout for quantitative dose verification, which requires amongst others a reliable calibration curve made from a series of polymarization gel dosimeters of identical geometry having been irradiated with different doses.\\
%
%
\section{Conclusions}
We present a prototype of a head shaped multimodality phantom for patient specific ``end-to-end'' tests including CT and MR imaging as well as a target volume providing range verification via gel dosimetry. We believe such phantoms to be increasingly necessary in the future due to the higher requirements for quality assurance in radiotherapy. \\
The phantom was found to be able to follow the entire workflow of an ion radiation. CT and MR contrast matched tissue values. Furthermore, the polimerization gel provided range verification with MRI. This phantom was a prototype study towards patient specific ``end-to-end'' tests in radiotherapy. Further developments should focus on more realistic anatomical matrices and more differentiated materials for further exploiting the excellent soft tissue contrast offered by MRI. Additionally, range verification might be extended to actual dose verification.\\
\section*{Acknowledgements}
The authors are grateful to Lars M\"{u}ller (DKFZ, E020), Navid Chaudhri (HIT), Martina Jochim, (DKFZ, E010) and the workshop at DKFZ for the assistance during measurements, data evaluation and production of the phantom. The authors should also like to thank Andreas Berg (MR-Center of Excellence, Vienna) for his valuable assistance with polymerization gel dosimetry.
\newpage
\section*{References}

\bibliography{PaperRef}

\newpage

\end{document}